\documentclass[pre,twocolumn,a4paper]{revtex4-1}
\usepackage{graphics}
\usepackage{amssymb}
\usepackage{amsmath}
\usepackage{wasysym}
\usepackage{latexsym}
\usepackage{graphicx}
\usepackage{epsfig}
\usepackage{subfigure}
\usepackage{float}

\begin{document}

\title{Influence of technological progress and  renewability on the sustainability of ecosystem engineers populations}

\author{Guilherme M. Lopes  and   Jos\'e F. Fontanari }     
\affiliation{Instituto de F\'{\i}sica de S\~ao Carlos,
  Universidade de S\~ao Paulo,
  Caixa Postal 369, 13560-970 S\~ao Carlos, S\~ao Paulo, Brazil}

\begin{abstract}
 Overpopulation and environmental degradation  due to  inadequate resource-use are outcomes of human's ecosystem engineering  that has  profoundly modified the world's landscape. Despite the age-old concern that unchecked  population and economic growth  may be   unsustainable, the prospect of  societal collapse remains contentious today. Contrasting  with  the usual  approach to  modeling human-nature interactions, which are based on  the Lotka-Volterra predator-prey model with humans as the predators and nature as the prey,  here we address this issue using a discrete-time population dynamics model of ecosystem engineers. The growth of the population of  engineers is modeled by the Beverton-Holt  equation with a density-dependent carrying capacity that is proportional to  the number of usable habitats. These habitats (e.g., farms) are the  products of the work of the individuals on the virgin habitats (e.g., native forests), hence the denomination  engineers of ecosystems to those agents. The human-made habitats  decay into  degraded habitats, which eventually regenerate into virgin habitats.  For slow regeneration resources, we find that  the dynamics is dominated by cycles of prosperity and collapse, in which the population reaches vanishing small densities. However,  increase of the efficiency of  the engineers to explore the resources eliminates the dangerous  cyclical patterns of feast and famine and leads to a stable equilibrium that balances  population growth and   resource availability.
  This finding supports the viewpoint of growth optimists that
technological progress may avoid collapse. 
\end{abstract}

\maketitle

\section{Introduction}

The gloomy consequences of rapid population growth and environmental degradation have been a major public concern,  probably first expressed explicitly  in the work of Thomas Malthus \cite{Malthus_98}, who believed that informed public policies could mitigate or even avoid altogether the happening of those circumstances. Following this line,  in the 1970s  the Club of Rome commissioned  a major enterprise  aiming at  modeling the  interactions between the earth and human systems using massive computer simulations capable of producing detailed  near-future scenarios \cite{Meadows_72,Meadows_04}.  Although efforts in that direction are still ongoing \cite{Randers_12}, this approach is somewhat abstruse as the complexity of the simulated models rivals that of the real-world, thus making it difficult to untangle the  effects of the model parameters.
An alternative approach to address such concerns, which we adopt here, is the study of minimal   models that link the population dynamics and the   resource dynamics. Works in this line typically build on Lotka-Volterra predator-prey models \cite{Murray_03}, where man is seen  as the predator and the resource base as the prey \cite{Brander_98,Motesharrei_14}.

Yet the predator-prey approach to human-nature dynamics misses the important fact that humans do not primarily prey on nature  but modify  it. In fact,   humans are the ultimate ecosystem engineers, who have  shaped
the world  into a more hospitable place for themselves,  most often with unlooked-for long term effects \cite{Smith_07}.  One such effect is
the collapse of  human societies likely caused by habitat destruction and overpopulation \cite{Tainter_90,Diamond_05}, hence the public concern on 
this matter. Accordingly, here we argue that the   baseline population dynamics model that is more suitable to address human-nature interactions is the ecosystem engineers model \cite{Jones_94,Gurney_96}, rather than the predator-prey model. The feature of the population dynamics of ecosystem engineers that makes this approach well suited to model those interactions   is that the growth of the  population is determined by the availability of usable habitats, which in turn are created by the engineers through the modification, and consequent  destruction, of virgin habitats. From a mathematical perspective, this feedback  results in a density-dependent carrying capacity that produces a very rich dynamic  behavior both in the continuous-time \cite{Gurney_96} and the discrete-time formulations \cite{Franco_17}.

More pointedly, the Gurney and Lawton model  for the population dynamics of ecosystem engineers describes the time evolution of   the density of engineers and of the  habitats, which can  exist in three different forms: virgin, usable  and degraded.  The transition of a habitat from  the virgin form  to  the usable form  occurs  due  to the  work of the engineers.  The usable habitats then degrade and eventually recover to become virgin habitats again.  Virgin and degraded habitats are unsuitable for the growth of the population. To expedite the numerical analysis, here we use a discrete-time approach where the logistic growth equation of the original model is replaced by the Beverton-Holt equation \cite{BH_57}. The weak nonlinearity of this equation prevents the  large  fluctuations on the population size  produced by the  Ricker equation \cite{Ricker_54}, which  may result in a   spurious chaotic dynamics \cite{Franco_17}.

We find that the survival of the population depends on two independent factors, viz., the competence of the engineers to transform natural resource base (e.g., native forests)  into useful goods (e.g., farms) and the population growth rate per generation. Survival is guaranteed provided that the product of these two factors is greater than the decay rate per generation of the usable habitats. However, survival is not without adversities in the case  the regeneration rate per generation of the degraded habitats is too  slow. In this case,  the dynamics exhibits cycles of prosperity and collapse, in which the population reaches vanishing small densities. Extinction does not occur because the fixed-point associated to it is unstable.  The regime of cycles is favored by large population growth rates. 

Most interestingly, we find that increase of the efficiency of  the engineers to explore the resource base, thus modeling  a high-tech society scenario  where a few individuals can extract and produce the goods needed for survival, eliminates the dangerous  cyclical patterns of feast and famine and leads to a smooth adjustment  between population and resources. This finding supports the viewpoint of growth optimists that
technological progress may avoid collapse.

In the case the resources are non-renewable, the long-term outcome is a doomsday scenario in which the ecosystem is reduced to    degraded habitats only. Since the population is not immediately affected by the disappearance of the virgin habitats, we can interpret the usable habitats as  a kind of wealth, i.e., surpluses that humans accumulate and that may extend their survival well beyond the point where the  natural resources are exhausted. This observation  offers  a  connection between ecosystem engineers models and economic models  of human-nature interactions \cite{Motesharrei_14}.

The rest of the paper is organized as follows. In Section \ref{sec:model}, we introduce the  discrete time version of Gurney and Lawton model of ecosystem engineers and derive the recursion equations for the  density of engineers as well as for the fractions of the three forms of habitats. In that section we  also derive and begin the analysis of   the local stability of the fixed point solutions of those equations. 
Section \ref{sec:res} presents the results of the numerical study of the eigenvalues of the Jacobian matrices that determine the local stability 
of the fixed points  as well as the results of the numerical iteration of the recursion equations. Finally, Section \ref{sec:disc} is reserved for our concluding remarks.

\section{Discrete-time population dynamics }\label{sec:model}

The  population dynamics of ecosystem engineering was   modeled originally by Gurney and Lawton in the 1990s using a continuous-time model  \cite{Gurney_96}, while discrete-time versions were considered only much more recently  \cite{Franco_17,Fontanari_18}. We recall that the main advantage of discrete-time formulations is that  they can easily be extended to incorporate a spatial dependence of the dynamic variables \cite{Kaneko_92}, as well as of the model parameters,  as neatly demonstrated in the classic studies of the spatial dynamics of host-parasitoid systems \cite{Hassell_91,Comins_92} (see \cite{Rodrigues_11,Mistro_12} for more recent contributions). However, discrete-time nonlinear systems usually exhibit chaotic  behavior, which is somewhat problematic from a biological perspective  \cite{Berryman_89} and are not observed in their continuous-time counterparts \cite{Sprott_03}. Here we show that this drawback can be circumvented  by a
suitable choice of the  equation that determines the growth of the population. In particular, whereas in previous studies the logistic equation of the continuous-time Gurney and Lawton model was replaced by the highly nonlinear Ricker equation \cite{Ricker_54}, here we use the  growth equation of Beverton and Holt \cite{BH_57}, whose much weaker  nonlinearity greatly  limits the  between-generations fluctuations and produces a stauncher representation of the continuous-time model.
 
We  assume that the population of engineers at generation $t$ is composed of $E_t$ individuals and that the carrying capacity of the environment depends on the  number of units of usable habitats available at generation $t$, which we denote by $H_t$. Thus, using the Beverton-Holt  equation we write the expected number of engineers at generation $t+1$ as  
\begin{equation}\label{EE1}
E_{t+1} = r E_t/  \left ( 1 + E_t/H_t \right ) 
\end{equation}
where  $r > 1$ is the intrinsic growth rate  of  the population of engineers  and the time-dependent carrying capacity is  $\left ( r - 1 \right ) H_t$. 
This model describes a system of  ecosystem engineers due to the  assumption that the usable habitats are produced by engineers working on virgin habitats  \cite{Gurney_96}. More pointedly, assuming that  there are  $V_t$ units of virgin habitats at generation $t$, a fraction $C \left ( E_t \right ) V_t$ of them will be  transformed in usable habitats by the engineers at the next generation $t+1$. Here  $C \left ( E_t \right )$ is any function that satisfies $ 0 \leq C \left ( E_t \right )  \leq 1$ for all $E_t$ and $C \left ( 0 \right ) = 0$. In other words, if there are no engineers at generation $t$, then  virgin habitats cannot  be transformed in usable ones. Hence  $ C \left ( E_t \right ) $ measures the efficiency of the engineer population to build usable habitats from the raw materials provided by the  virgin habitats.  

Usable habitats  decay into degraded habitats that are useless to the engineers, in the sense that they are inhabitable and  lack the raw materials needed to build usable habitats.  Denoting by $ \delta \in \left [ 0, 1 \right ]$ the  probability that a unit of usable habitat decays and becomes  a unit of degraded habitat   in one generation, we  write  the expected number of units of usable habitats at generation $t+1$ as
\begin{equation}\label{HH1}
H_{t+1} = \left ( 1 - \delta \right ) H_t + C \left ( E_t \right ) V_t.
\end{equation}
In this paper we assume that  the decay probability $\delta$  is a density independent parameter. We recall
 that in the  Gurney and Lawton model the resources are represented by the  virgin habitats, whose  probability of change into usable habitats   is density dependent, i.e.  $C = C \left ( E_t \right )$.  For example, in an island scenario, the virgin habitats can be thought of as the native forests whereas the usable habitats are  the lands cleared for crops,   whose degradation, due mainly to erosion and soil depletion of nutrients, is more suitably modeled by a constant decay  probability $\delta$, rather than by a density-dependent one. (i.e., $\delta = \delta \left ( E_t \right ) $). However, in an economic context where  the habitats represent, for instance, oil wells,  the decay probability should be modeled by a density-dependent  parameter, of course.

To take into account the possibility that the degraded habitats will eventually recover and become virgin habitats again, we introduce the parameter $ \rho \in \left [ 0, 1 \right ]$ that gives the probability that one unit of degraded habitat becomes one unit of virgin habitat in one generation. Thus, denoting the units of degraded habitats at generation $t$ by $D_t$ we  write 
\begin{equation}\label{DD1}
D_{t+1} = \left ( 1 - \rho \right ) D_t + \delta H_t .
\end{equation}
Thus $\rho$ is the regeneration rate per generation that measures the renewability of the resource base.
Finally, recalling that virgin habitats are destroyed by the action of the engineers and created by the restoration  of the degraded habitats, we write the recursion equation for the expected number of units of virgin habitats as
\begin{equation}\label{VV1}
V_{t+1} =    \left [ 1 - C \left ( E_t \right ) \right ]  V_t +  \rho  D_t ,
\end{equation}
 which completes the description of the feedback  interactions between habitats and engineers.
 
Since habitats   can only be transformed  from one form to another, 
the  total number of units of habitats 
  $ V_{t+1} + H_{t+1}  + D_{t+1} = V_{t} + H_{t}  + D_{t}   = T$ is conserved.  This fact allows us to introduce
 the habitat fractions $v_t \equiv V_t/T$, $h_t \equiv H_t/T$ and $d_t \equiv D_t/T$ that satisfy $v_t + h_t + d_t =1$ for all $t$. 
 Accordingly, we define also the density of engineers $e_t = E_t/T$ that, differently from the habitat fractions, may take on values greater than 1. In terms of these  intensive quantities, the above recursion equations are rewritten as
\begin{eqnarray}
e_{t+1} &  =  & r e_t /\left ( 1 + e_t/h_t \right )  \label{e} \\
h_{t+1} & = & \left ( 1 - \delta \right ) h_t + c \left ( e_t \right ) v_t  \label{h} \\
v_{t+1} & = &  \rho \left ( 1 - v_t -h_t  \right ) + \left [ 1- c \left ( e_t \right ) \right ] v_t, \label{v}
\end{eqnarray}
 where we have used $ d_t = 1 - v_t - h_t$ and $c \left ( e_t \right ) \equiv C \left ( T e_t  \right )$.

To complete the model we must specify  the density-dependent probability  $c \left ( e_t \right )$, which measures the engineers' efficiency to transform the virgin habitats into usable ones. The function $c \left ( e_t \right )$ incorporates  the cooperation  strategies  used  to build  structures (e.g., beaver dams, termite mounds and  anthills) to respond to external and internal threats to the population survival \cite{Francisco_16,Reia_17}. For humans,  this function  incorporates the beneficial effects (from their  perspective) of the technological advancements   that enable a more efficient harvesting of natural resources \cite{Basalla_89}.  Here we  consider the continuous function
\begin{equation}\label{c}
c \left (e_t \right )=\left \{
			\begin{array}{c l}
                     \alpha  e_t  +  \left ( 1 - \alpha \right ) e_t^2 & \mbox{if $e_t < 1 $}. \\
                     1 &  \mbox{if $e_t \geq 1 $}.
            \end{array}
            		\right.
 \end{equation}
%
%
%\begin{equation}\label{c}
 %c \left ( e_t \right ) = 1 - \exp \left ( - \alpha e_t \right ),
%\end{equation}
%
 where $\alpha \in \left [ 0, 1 \right ]$ is the efficiency  parameter that measures the competence of the engineers in transforming  natural resources into useful goods.  The range of variation of $\alpha$ guarantees that $c \left (e_t \right ) \leq 1$ regardless of the value of $e_t$.
For $\alpha \ll 1$ we have a scenario where  the efficient exploration of the virgin habitats requires a high density of engineers. In this  low-tech scenario, we expect to observe an Allee effect, in which the population quickly goes extinct below a  critical density \cite{Courchamp_08}.  
We note that when $ e_t > 1$  all the available virgin habitats are transformed into usable ones in just  a single generation. This arbitrary threshold value does not produce any significant effect on our results  since we find that  $e_t < 1$ at all times in the parameter regions of interest, i.e., in the regions close to the transitions between the fixed point attractors  and the limit cycles.

\subsection{Fixed-point solutions}

In this section we study  the fixed-point solutions of the system of  recursion equations (\ref{e})-(\ref{v}) that are obtained by setting $e_{t+1} = e_t = e^*$, $h_{t+1} = h_t = h^*$ and $v_{t+1} = v_t = v^*$. Next we consider separately the two cases, $e^* =0$ (zero-engineers fixed point) and $e^* > 0$ (finite-engineers fixed point). The local stability of the fixed point solutions are determined by the eigenvalues of the Jacobian matrix of the system of recursion equations (\ref{e}), (\ref{h}), (\ref{v}),
\begin{equation}\label{Je}
J_{e_t} = 
\begin{bmatrix}
 r/  \left ( 1 + x_t \right )^2   &~~~~r \left [ x_t/  \left ( 1 + x_t \right ) \right ]^2  &~~~~ 0  \\
   v_t  c' \left ( e_t \right )  &~~~~  1-\delta &~~~~ c \left ( e_t \right ) \\
   - v_t  c' \left ( e_t \right ) &~~~~  -\rho  &~~~~ 1 -\rho  - c \left ( e_t \right )  
  \end{bmatrix},
 \end{equation} 
where $x_t = e_t/h_t$ and $ c' \left ( e_t \right ) = d c\left ( e_t \right )/d e_t$.

\subsubsection{Zero-engineers fixed point}\label{sec:zero}

In the absence of the  engineers (i.e., $ e^* = 0$)  all usable habitats will degrade and  then recover to virgin habitats (provided that $\rho > 0$), so we have $h^* = 0$ and $v^* = 1$. To circumvent the indetermination of the ratio $e^*/h^*$, we write an equation for $x_t$ dividing eq.\ (\ref{e}) by eq.\  (\ref{h}),
\begin{equation}\label{x2}
x_{t+1}   =  \frac{ r x_t }{ \left ( 1 + x_t \right )  \left [ 1 - \delta  +  x_t  v_t  \hat{c} \left ( e_t\right ) \right ]},
\end{equation}
where $\hat{c} \left ( e_t\right ) = c \left ( e_t \right )/e_t$. Thus the fixed point $x_{t+1}=x_t = x^* > 0$  is given by the positive solution of the quadratic equation
\begin{equation}\label{x3}
 \left ( 1 + x^* \right )   \left [ 1 - \delta  +  \alpha x^*   \right ] =  r , 
\end{equation}
 where we  have used $  \hat{c} \left ( 0 \right ) = \lim_{e^* \to 0}  c \left ( e^* \right )/e^* = c' \left ( 0 \right ) = \alpha $.
The quadratic equation (\ref{x3}) has two real roots, one positive and the other negative, for all values of the model parameters. In particular, 
 for $\delta = \alpha \left ( r - 1\right )$ the solution is $x^* = r-1$ and we can easily show that $x^* > r-1$ for $ \delta > \alpha \left ( r - 1\right ) $ and $x^* < r-1$ for $ \delta < \alpha \left ( r - 1\right )$, a result that  will proven  helpful to verify  the local stability of this fixed point.
      
In fact, the local stability of the zero-engineers fixed point  is 
determined by requiring that the three eigenvalues of the Jacobian matrix 
\begin{equation}\label{J0}
J_{e^* =0} = 
\begin{bmatrix}
 r/  \left ( 1 + x^* \right )^2   &~~~~ r \left [ x^*/  \left ( 1 + x^* \right ) \right ]^2   &~~~~ 0  \\
   \alpha  &~~~~  1-\delta &~~~~ 0  \\
  -\alpha &~~~~  -\rho  &~~~~ 1 -\rho   
  \end{bmatrix}
 \end{equation} 
have norms less than 1.   The eigenvalues  of the Jacobian matrix (\ref{J0}) are  $\lambda_1 = r/\left ( 1 + x^* \right )  $, 
$\lambda_2 = r/\left ( 1 + x^* \right )^2 - \alpha x^* $ and  $\lambda_3 = 1 - \rho$.
The condition $ \mid \lambda_1  \mid < 1$  is satisfied provided  that $x^* > r - 1$, which is guaranteed for  $\delta > \alpha \left ( r - 1\right )$, whereas the condition $ \mid \lambda_3  \mid < 1$ is always satisfied  since $\rho < 1$. 
The proof that the condition $ \mid \lambda_2  \mid < 1$ is always satisfied whenever $\lambda_1 < 1$ is very simple. Since $\alpha> 0$ and $x^* > 0$ we can write $ \lambda_2  < r/\left ( 1 + x^* \right )^2 = \lambda_1/\left ( 1 + x^* \right ) < 1/\left ( 1 + x^* \right ) < 1$. Next we need  to show that  $ \lambda_2 > -1$. To do so  we use eq.\ (\ref{x3}) to write $\alpha x^* = \lambda_1 - 1 + \delta$ so that $\lambda_2 = 1 - \delta - x^* \lambda_1/\left ( 1 + x^* \right ) >  1 - \delta -1 > -\delta > -1$. 

We note that this stability analysis describes the convergence (or divergence) of the system of recursion equations for $e_t$, $h_t$ and $v_t$, eqs.\ (\ref{e}), (\ref{h}) and  (\ref{v}), respectively,  to the fixed point $e^*=h^* =0$ and $v^* =1$ and that equations (\ref{x2}) and (\ref{x3}) were used only to specify the ratio
$\lim_{t \to \infty} e_t/h_t$ that is required to evaluate the Jacobian (\ref{Je}). We mention this because  use  of recursion equations for
$e_t$, $x_t$ and $v_t$ instead, yields a different Jacobian for the zero-engineers fixed point, which exhibits the same   eigenvalues $\lambda_1$ and $\lambda_3$ as the Jacobian (\ref{J0}), but a different eigenvalue $\lambda_2$.  However, the local stability of the fixed point is again determined solely  by the norm of  the common eigenvalue $\lambda_1$ so this issue has no effect in  our findings regarding the zero-engineers fixed point.

In sum, the condition for the stability of the zero-engineers fixed point is $ \delta < \alpha \left ( r - 1\right )$, which, as expected, does not depend  on the regeneration rate $\rho $ since there is no shortage of virgin habitats (i.e., $v_t \approx 1$) in the vicinity of this fixed point.   

\subsubsection{Finite-engineers fixed point}\label{subsec:FE}

Since $e^* > 0$ we have   $h^* =e^*/ \left ( r -1  \right )$ and $v^* = 1 -  e^* \gamma/   \left ( r -1  \right )$ with $e^*$ given by the solution of the   equation 
\begin{equation}\label{e*1}
\delta e^* =  c \left ( e^* \right ) \left [  r -1   -   e^* \gamma \right ] ,
\end{equation}
where we have introduced the notation $\gamma = \left ( 1 + \delta/\rho \right )$. 

Let us assume  that $e^* < 1$ so we can  use eq.\ (\ref{c}) to rewrite  eq.\ (\ref{e*1}) as  the  quadratic equation
 \begin{equation}\label{e*2}
 \left ( 1 - \alpha \right ) \gamma (e^*)^2 - e^* \left [ \left ( 1 - \alpha \right )  \left ( r -1  \right ) - \alpha \gamma \right ] + \delta - \alpha  \left ( r -1  \right )= 0 .
 \end{equation}
 The local stability  of the two solutions of this equation is determined by the   the Jacobian matrix (\ref{Je}) that is rewritten as 
\begin{equation}\label{Je*}
J_{e^*<1} = 
\begin{bmatrix}
 1/r    &~~~~ \left ( r -1 \right)^2/r    &~~~~ 0  \\
   v^*  c' \left ( e^* \right )  &~~~~  1-\delta &~~~~ c \left ( e^* \right ) \\
   - v^*  c' \left ( e^* \right ) &~~~~  -\rho  &~~~~ 1 -\rho  - c \left ( e^* \right )  
  \end{bmatrix}.
 \end{equation} 
The main difficulty of the analysis of the eigenvalues of this Jacobian matrix is that two eigenvalues are complex, leading to  the enthralling dynamic behavior  exhibited by the  recursion equations (\ref{e}), (\ref{h}), (\ref{v}) with the prescription (\ref{c}). However, this problem can  easily  be solved numerically as follows. First we determine the  critical points of the characteristic cubic polynomial, which allows us  to obtain one real root of the characteristic equation with high numerical accuracy using the bisection method.  (We recall that the  critical points of a function are those values of the variable where the slope of the function is zero.) Then this  root is factored out  and the remaining roots of the characteristic equation are  found by solving a quadratic equation. Actually, in the case of complex roots, we do no need to solve this quadratic equation since
the local stability is determined by the norm of those roots that is given by the square root of the ratio between the constant and the quadratic coefficient. The fixed point solution is locally stable  provided that the norms of  all three roots of the characteristic polynomial are less than 1.
We find that the smaller root of the fixed-point equation (\ref{e*2}) is  always locally unstable.

Now we consider the case $e^* > 1$. Equation (\ref{e*1}) yields 
 \begin{equation}\label{e>1}
e^* =  \frac{ r -1 }{ 1 + \delta \left (1 + 1/\rho \right )}
\end{equation} 
that holds for $ \delta < \left ( r -2 \right)/ \left ( 1 + 1/\rho \right)$.
In this case, the Jacobian matrix (\ref{Je}) becomes
\begin{equation}\label{Je*1}
J_{e^*>1} = 
\begin{bmatrix}
 1/r    &~~~~ \left ( r -1 \right)^2/r    &~~~~ 0  \\
  0  &~~~~  1-\delta &~~~~ 1 \\
   0 &~~~~  -\rho  &~~~~  -\rho    
  \end{bmatrix}.
 \end{equation} 
Hence $\lambda_1 = 1/r < 1$ is always a root of the characteristic polynomial and the other two roots are given by the quadratic equation $g \left ( \lambda \right ) = \lambda^2 - \lambda \left ( 1 - \rho - \delta \right ) + \rho \delta = 0$. In the case of complex roots, we have $\mid \lambda \mid^2 = \rho \delta < 1$.  In the case of real roots,  we  use  that $g(1)> 0$ and $g(-1)>0$  and that  the critical point of $g(\lambda)$ is   between $-1$ and $1$ to guarantee that the absolute value of those roots are  less than 1. Hence the fixed point (\ref{e>1}) 
 is always locally stable.

\section{Results}\label{sec:res}

%----------------------------------------------------------------------------------
\begin{figure}
\begin{center}
\includegraphics[width=0.48\textwidth]{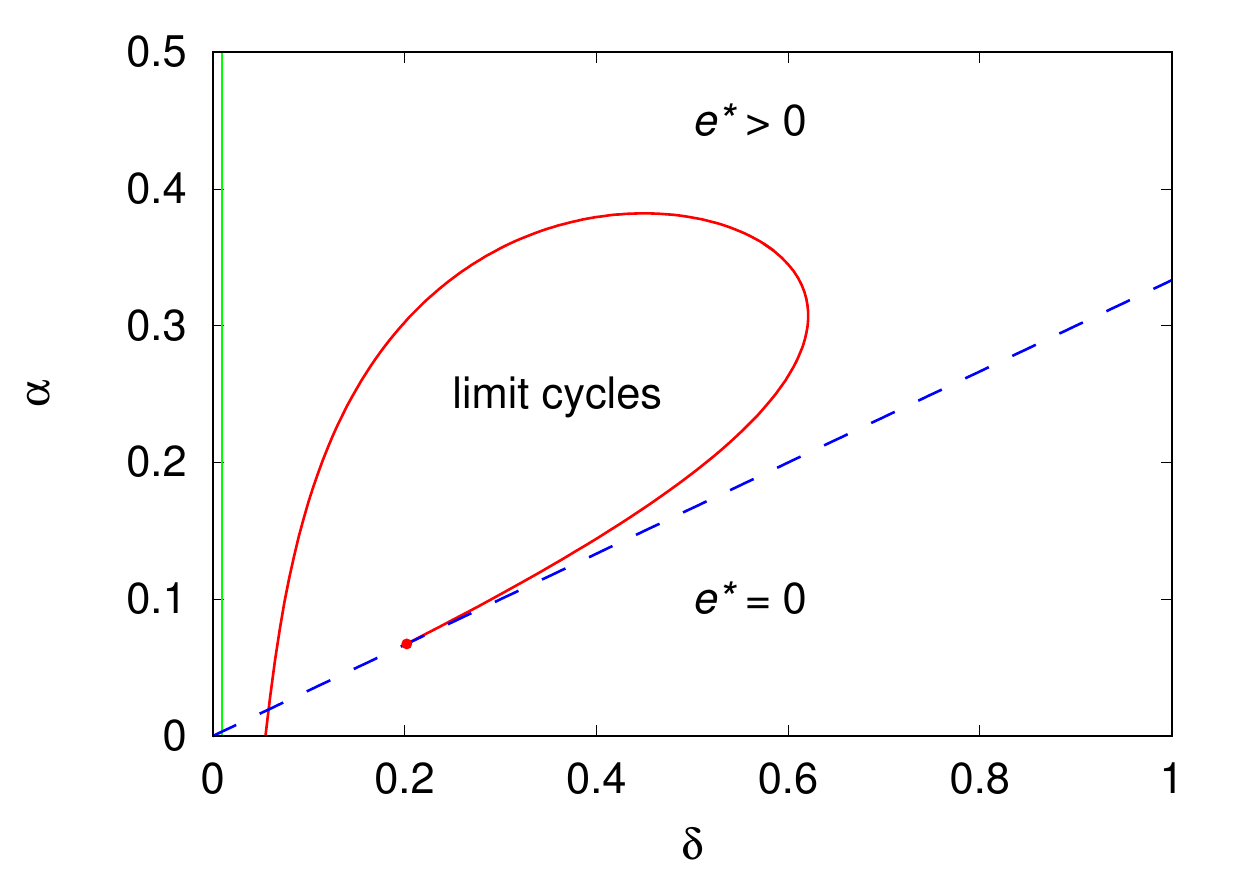}
\caption{Regions in the space of the parameters  $\left ( \delta, \alpha \right )$  where the fixed points  $e^*=0$ and $0< e^* < 1$ are locally stable for  $r=4$ and $\rho = 0.005$. The dashed blue line  $\alpha = \delta/3$ delimits the region of stability of the zero-engineers fixed point.  There is a region of bistability for $\alpha < \delta/3$ and $\delta < 0.055$. The limit cycles  appear only in the regions where the fixed points are unstable. The fixed point $e^* > 1$ exists and is stable only in the region to the left of the vertical  green line at  $\delta = 2/201 \approx 0.01$. The red bullet indicates the value of $\delta$ beyond which the transition over the dashed line is continuous.    }
\label{fig:1}
\end{center}
\end{figure}
%----------------------------------------------------------------------------------
%----------------------------------------------------------------------------------
\begin{figure}
\begin{center}
\includegraphics[width=0.48\textwidth]{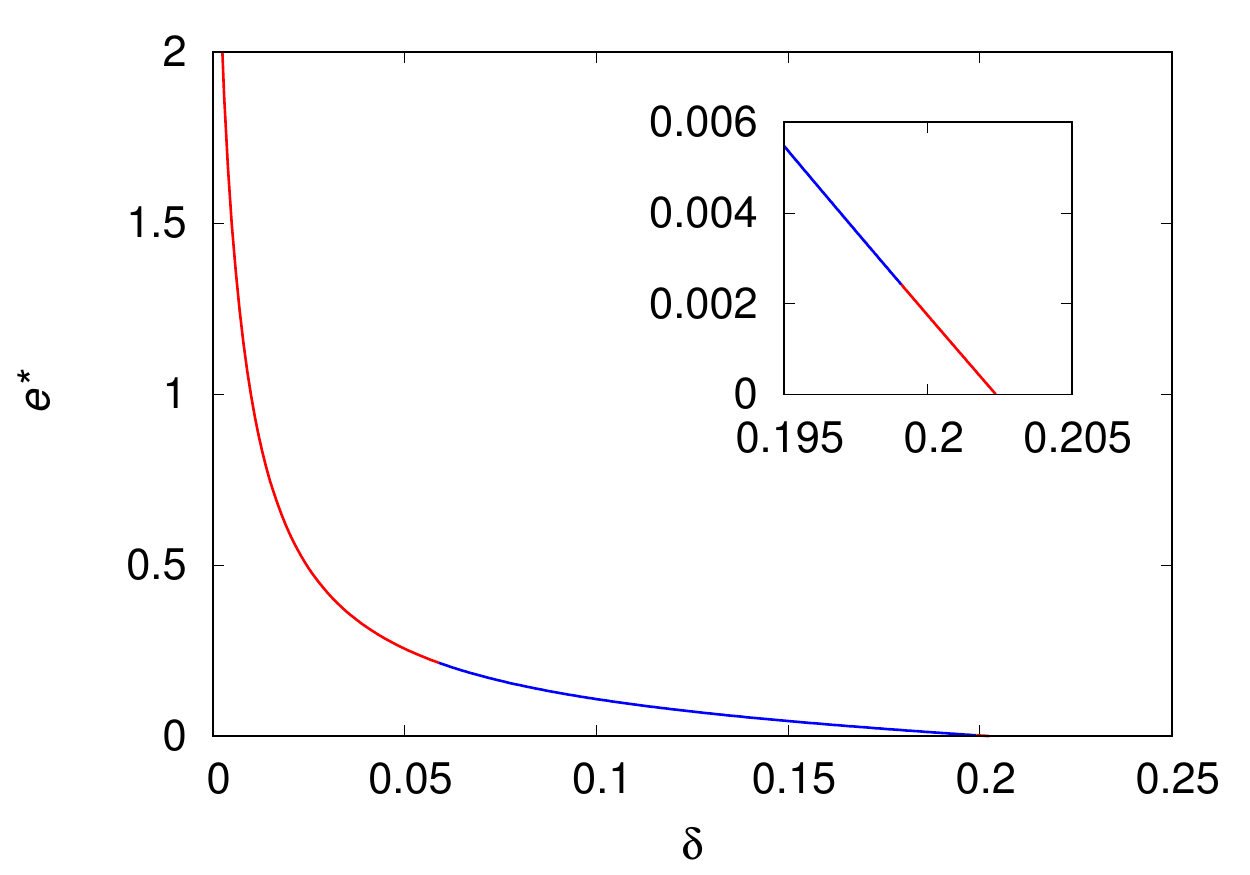}
\caption{Fixed-point values of the density of engineers $e^*$  as function of the decay probability $\delta$  at the transition line $\alpha = \delta/ \left ( r -1  \right )$ for  $r=4$ and  $\rho = 0.005$. The fixed point is stable in the red segment of the curve and unstable in the blue segment. The inset shows that the finite-engineers fixed point becomes stable  again close to the point where it vanishes, indicated by the red bullet in Fig.\ \ref{fig:1}.  }
\label{fig:2}
\end{center}
\end{figure}
%----------------------------------------------------------------------------------

The local stability of the  fixed points $e^* =0$ and $0 < e^* <1$ is determined by the requirement that the norms  of the  eigenvalues of the Jacobian matrices (\ref{J0}) and (\ref{Je*}), respectively,   are  less than 1. Accordingly, Fig.\  \ref{fig:1} shows the regions of stability of those fixed points. There is a large region in the space of parameters $\left ( \delta, \alpha \right )$ where the fixed points are unstable and the solutions of the recursion equations exhibit periodic behavior.  The drop-shaped region  delimits the regions of stability of the finite-engineers fixed point.

%----------------------------------------------------------------------------------
\begin{figure*}
\begin{center}
 \subfigure{\includegraphics[width=0.325\textwidth]{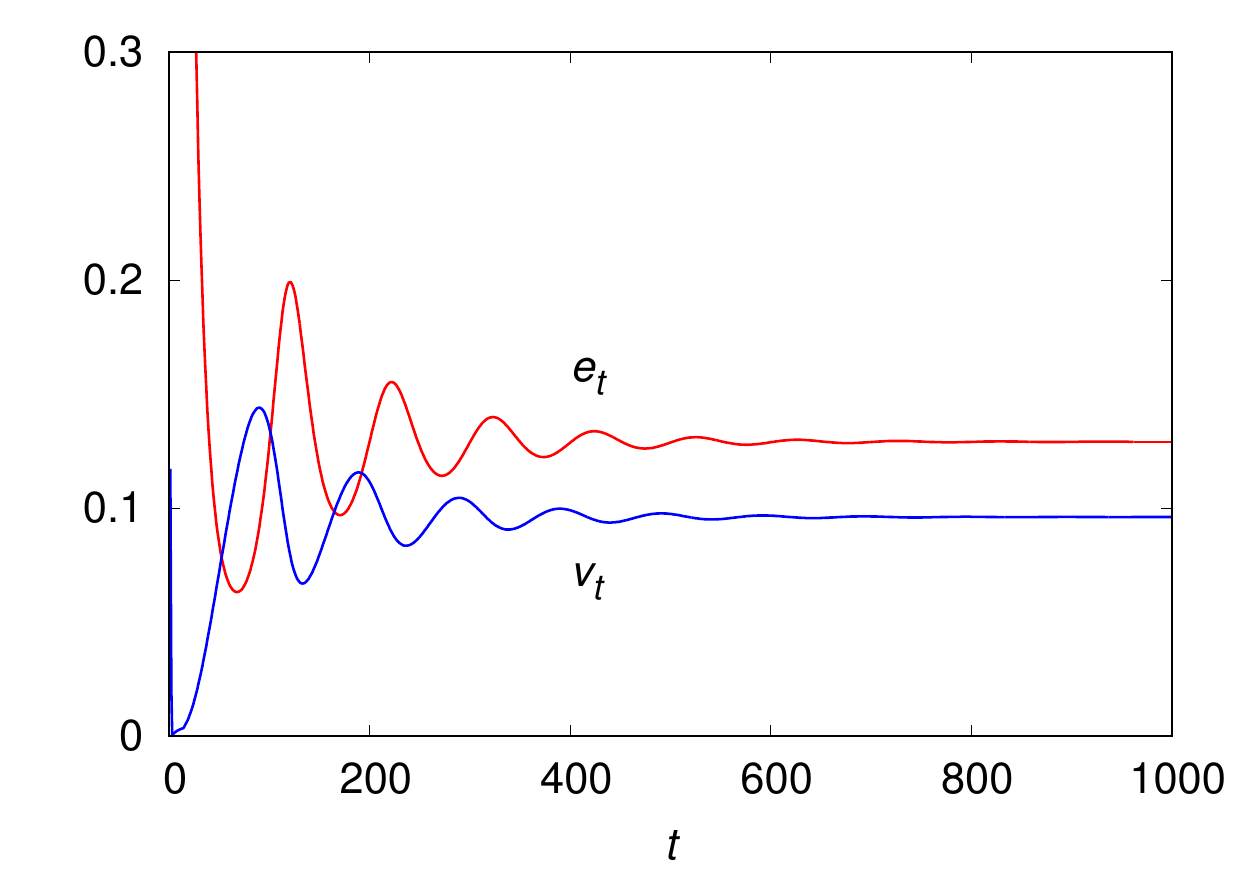}} 
 \subfigure{\includegraphics[width=0.325\textwidth]{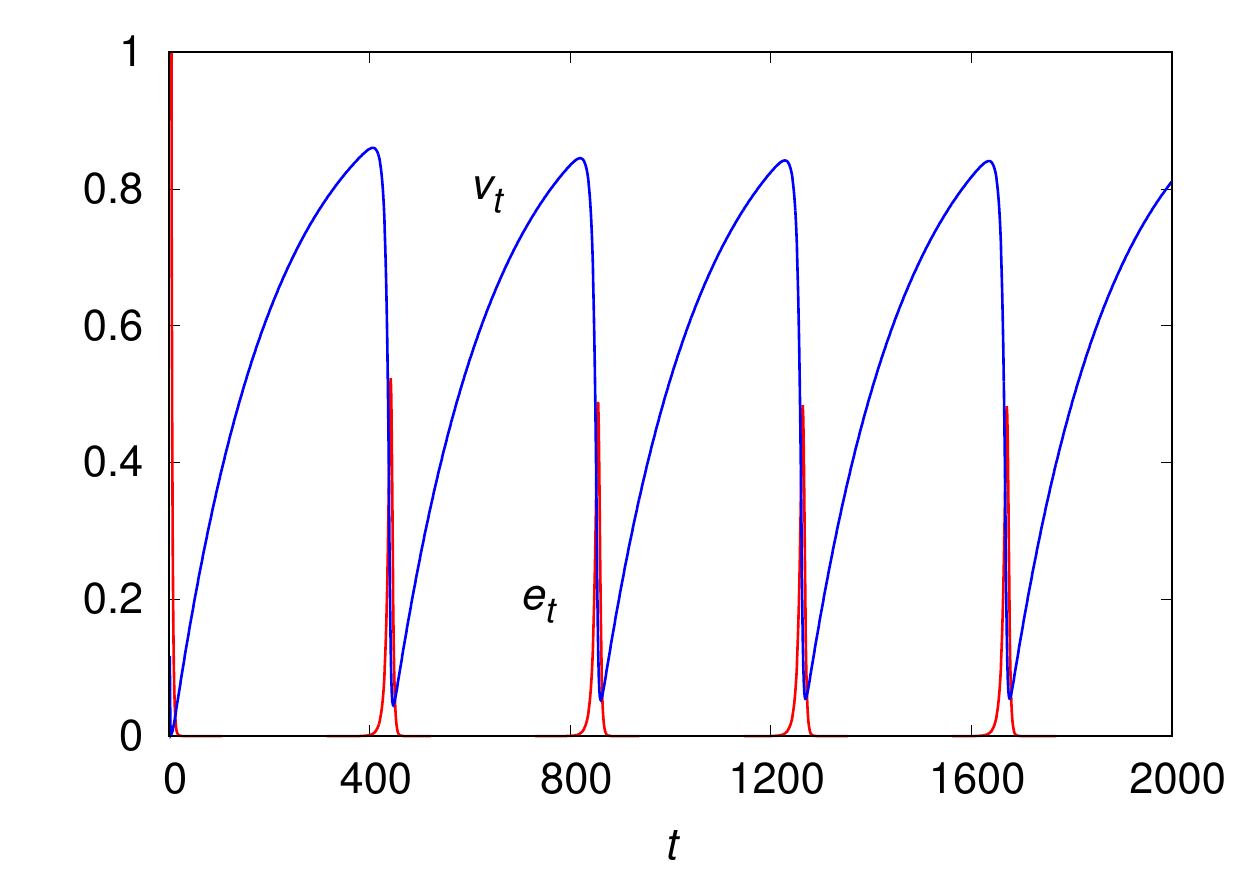}} 
 \subfigure{\includegraphics[width=0.325\textwidth]{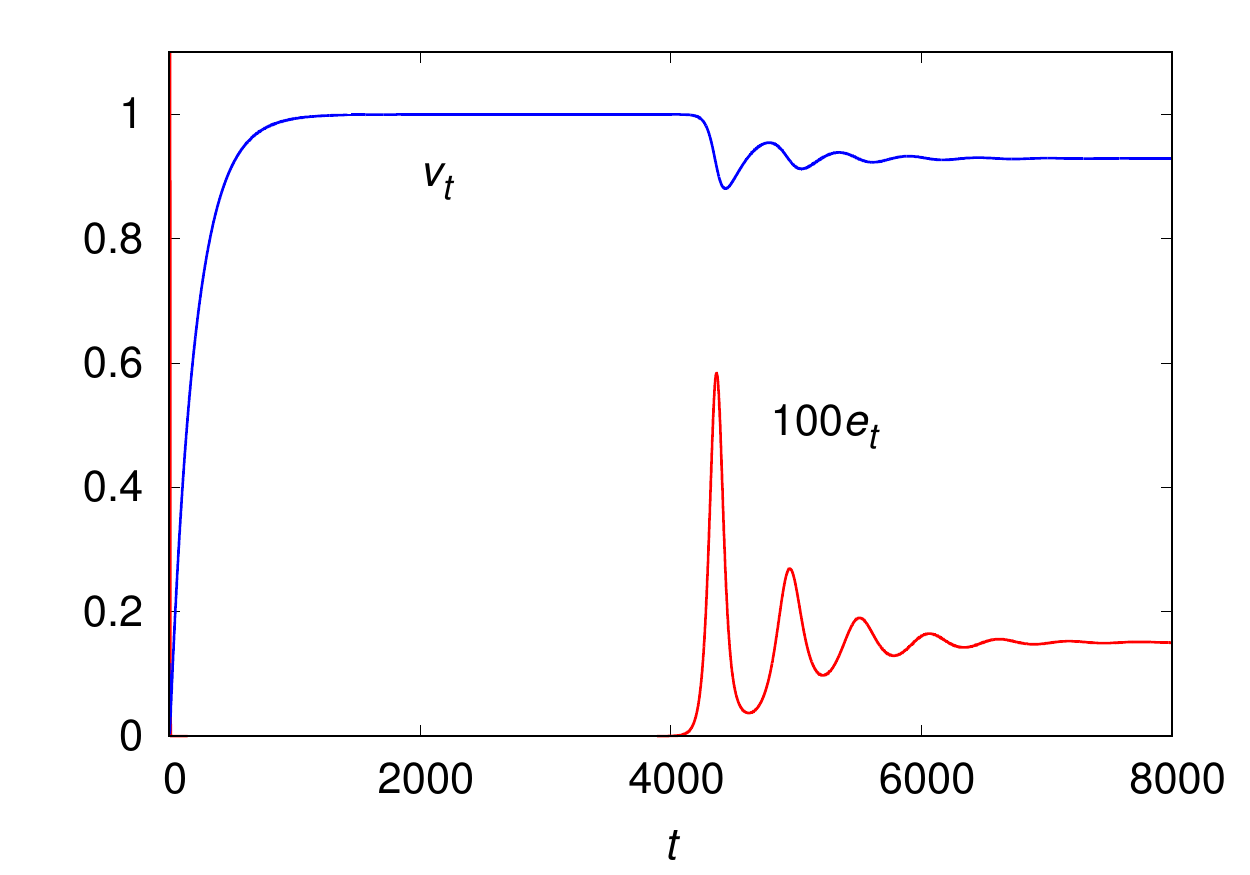}}
\caption{Density of engineers $e_t$ (red curves) and fraction of virgin habitats $v_t$ (blue curves)  for $r=4$, $\rho = 0.005$, $\alpha =0.25$ and $\delta = 0.1$ (left panel), $\delta = 0.4$ (middle panel) and $\delta = 0.7$ (right panel). In the right panel, the density of engineers is magnified by a factor 100 in order to be visible in the scale of the figure.
 }  
\label{fig:3}  
\end{center}
\end{figure*}
%----------------------------------------------------------------------------------

%----------------------------------------------------------------------------------
\begin{figure*}
\begin{center}
 \subfigure{\includegraphics[width=0.325\textwidth]{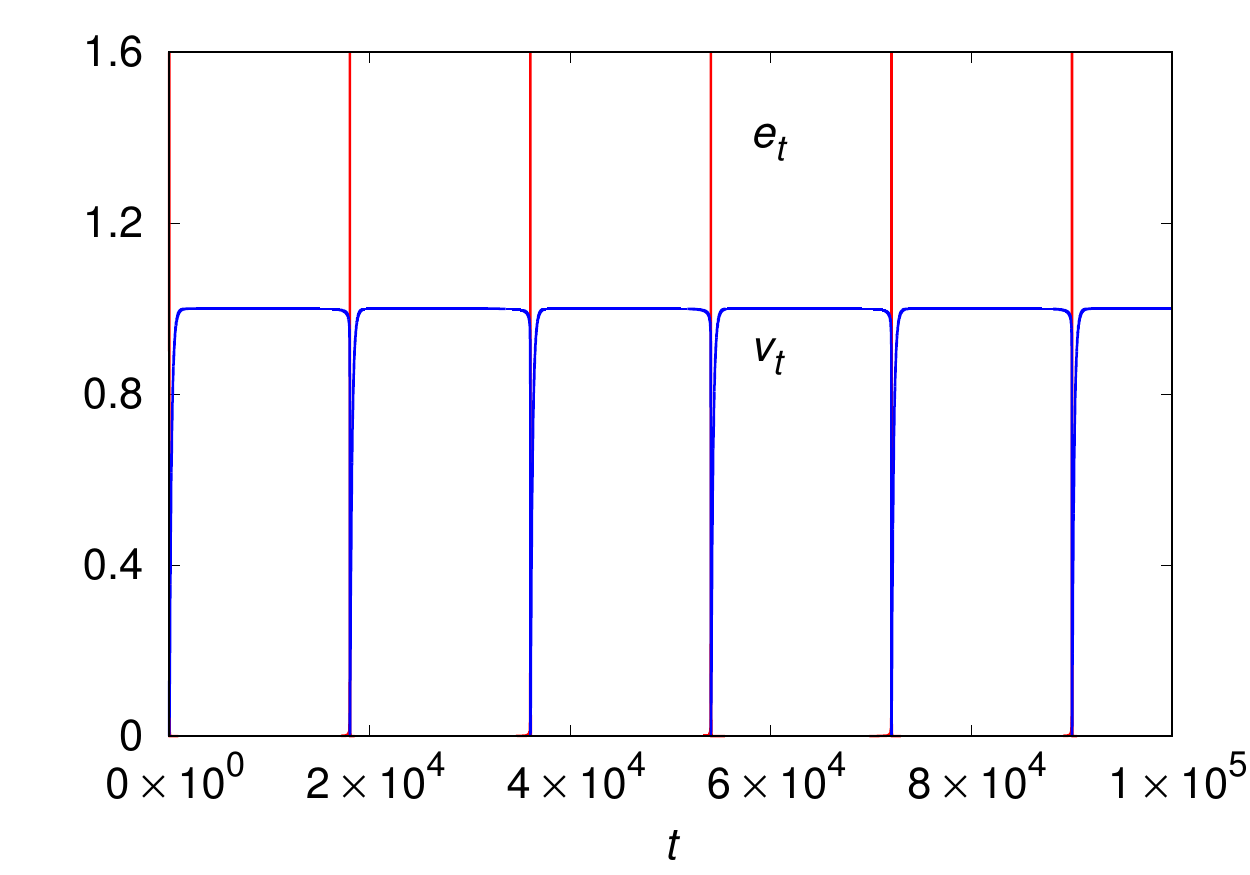}} 
 \subfigure{\includegraphics[width=0.325\textwidth]{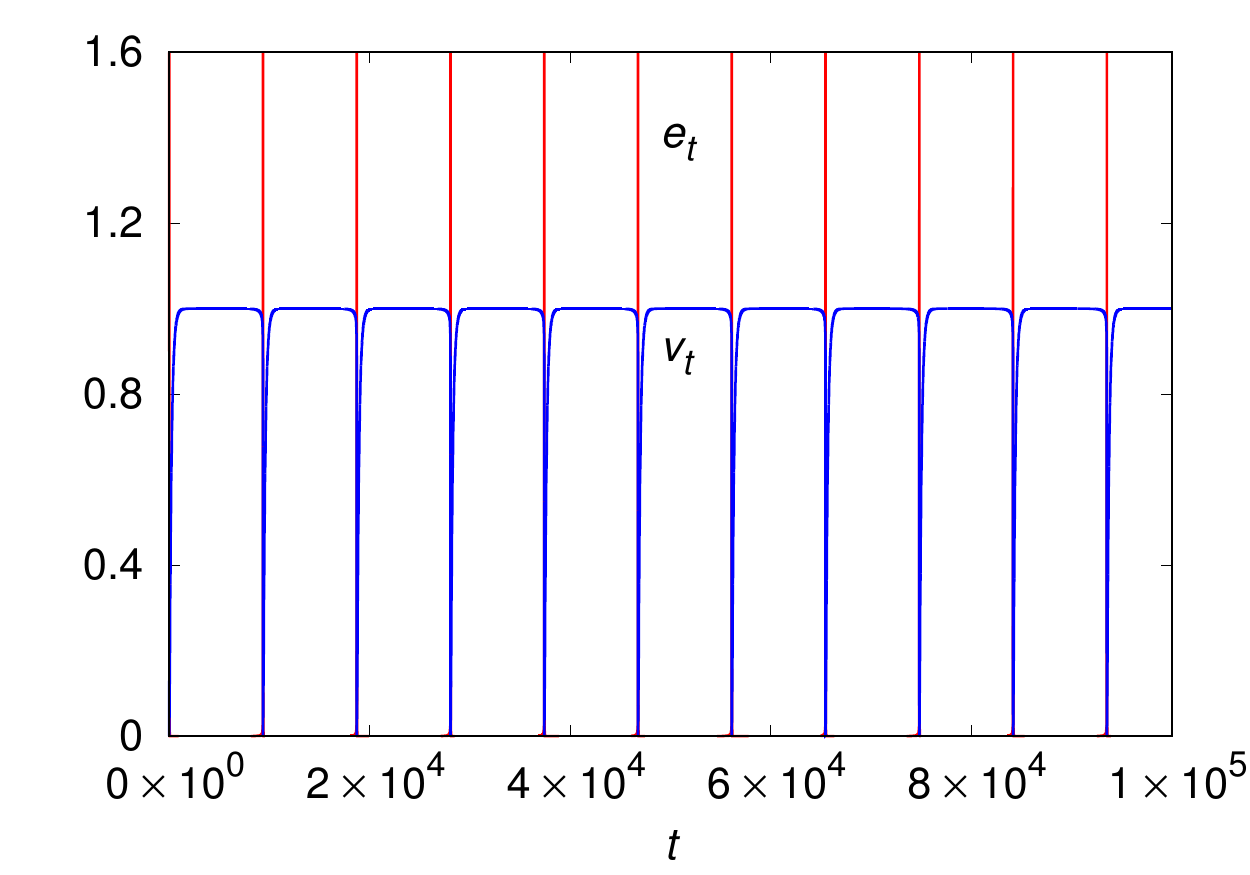}} 
 \subfigure{\includegraphics[width=0.325\textwidth]{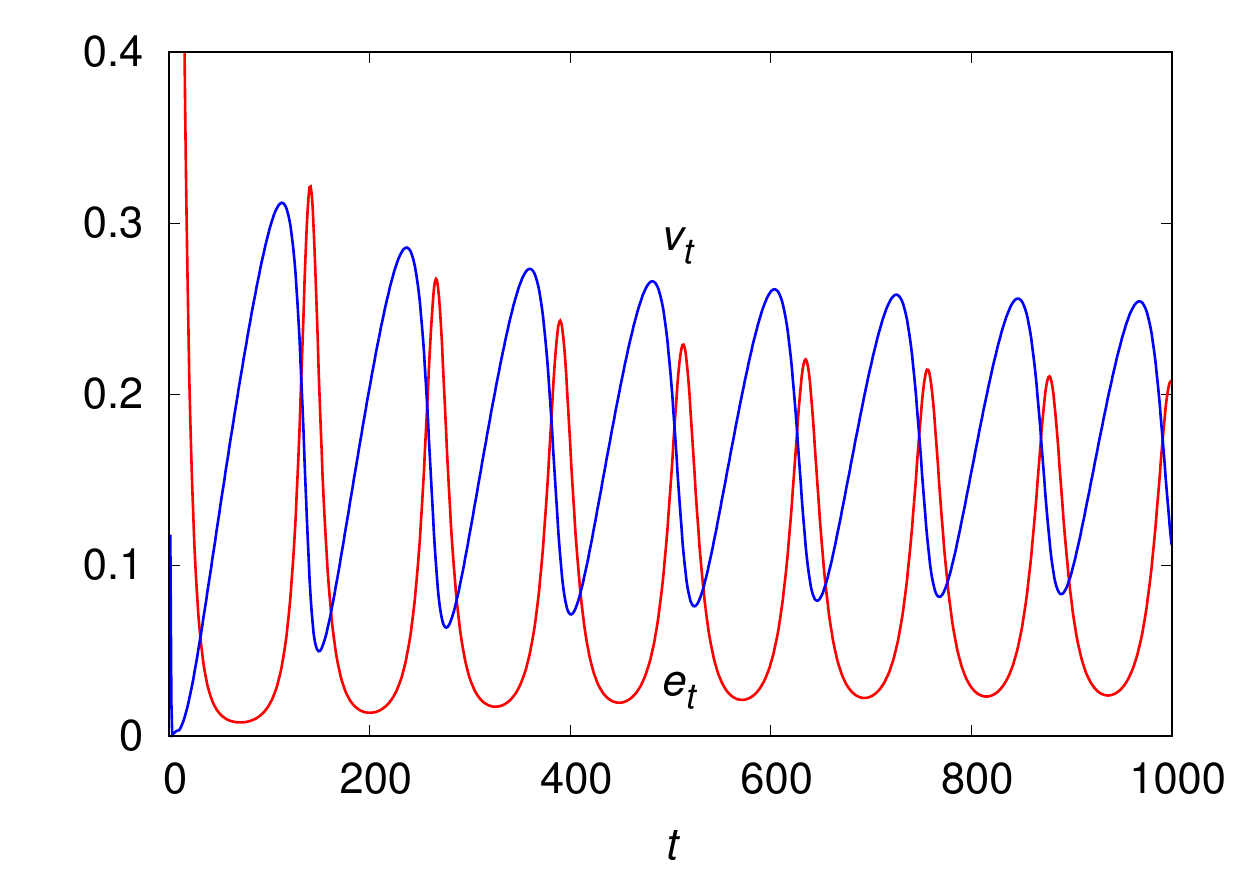}}
\caption{Density of engineers $e_t$ (red curves) and fraction of virgin habitats $v_t$ (blue curves)  for $r=4$, $\rho = 0.005$, $\delta =0.15$ and $\alpha = 0.0505$ (left panel), $\alpha= 0.051$ (middle panel) and $\alpha = 0.24$ (right panel).
 }  
\label{fig:4}  
\end{center}
\end{figure*}
%----------------------------------------------------------------------------------

To better understand the results shown in Fig.\  \ref{fig:1},  we first examine whether  eq. (\ref{e*2})  admits  a physical solution with a vanishingly  small density of engineers. In fact, for $e^* \ll 1$ we have
\begin{equation}\label{e*3}
e^* \approx \frac{\alpha  \left ( r -1  \right ) - \delta}{\alpha \gamma - \left ( 1 - \alpha \right )  \left ( r -1  \right )  } ,
\end{equation}
so that  the density of engineers vanishes  continuously as the line $\alpha = \delta/ \left ( r -1  \right )$   is approached from the region of large $\alpha$, provided the denominator in eq. (\ref{e*3}) is positive. i.e., $\alpha \gamma - \left ( 1 - \alpha \right )  \left ( r -1  \right ) > 0$.  In Fig.\ \ref{fig:1} we indicate with a red bullet the point  $\delta = \delta_c \approx 0.202$ and $\alpha = \alpha_c \approx 0.067$ at which both the numerator and the denominator  in eq. (\ref{e*3})  vanish, meaning  that $e^*=0$ is a double root of the quadratic equation (\ref{e*2}). Thus,  there is a continuous transition between the finite and the zero-engineers fixed point at $\alpha = \delta/ \left ( r -1  \right )$ for $\delta > \delta_c$.

%----------------------------------------------------------------------------------
\begin{figure}
\begin{center}
\includegraphics[width=0.48\textwidth]{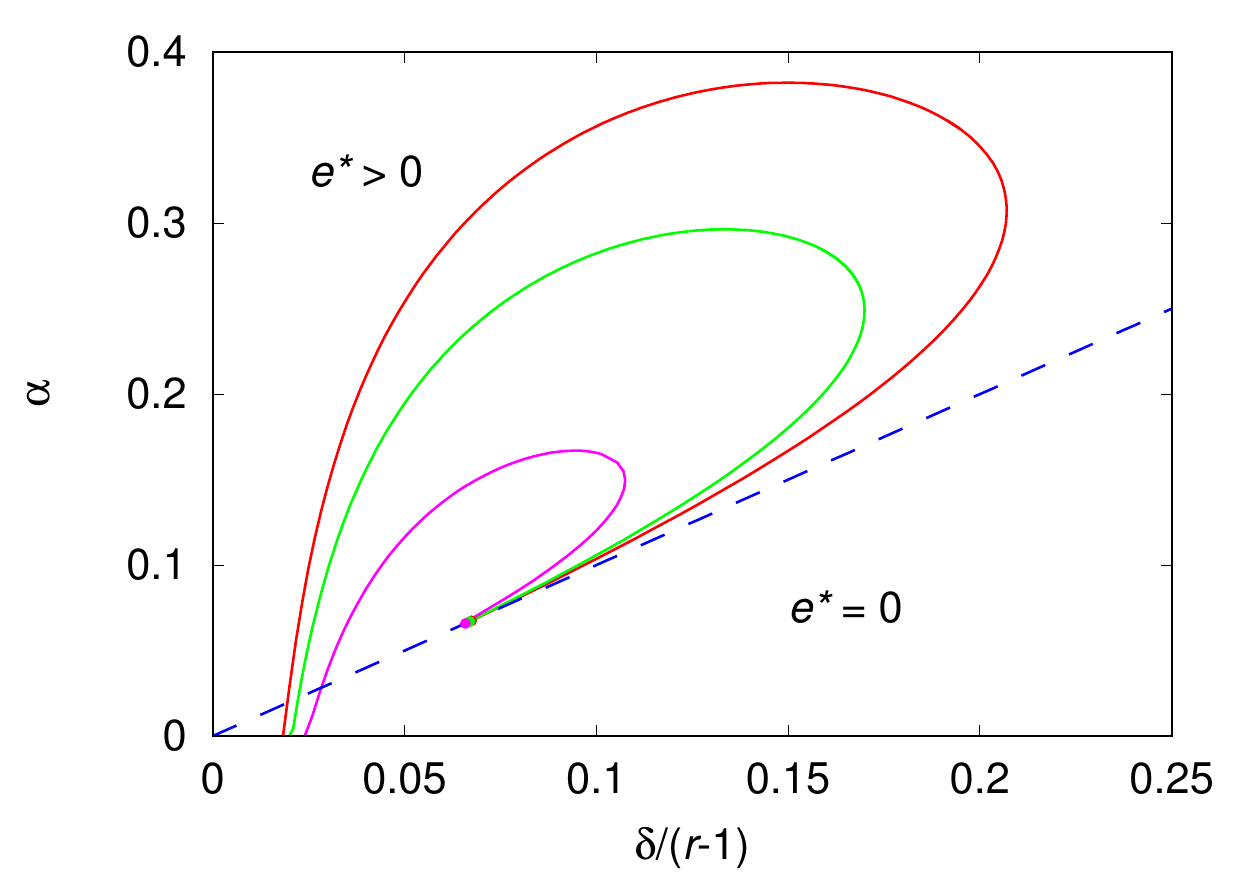}
\caption{Regions in the space of the parameters  $\left ( \delta, \alpha \right )$  where the  fixed points are locally stable  for   $\rho = 0.005$ and  $r=4$ (red curve),  $r=3$ (green curve) and $r=2$ (magenta curve). The conventions for the labelling of the regions  are the same as in  Fig.\ \ref{fig:1}, but we have omitted the regions where the fixed point  $e^* > 1$ appears.  The zero-engineers fixed point is stable below the dashed blue line $\alpha = \delta/\left ( r-1 \right )$.  }
\label{fig:5}
\end{center}
\end{figure}
%----------------------------------------------------------------------------------

Next, we focus on the fixed-point solutions at the  transition line $\alpha = \delta/ \left ( r -1  \right )$. In this case, eq. (\ref{e*2}) has the solution $e^* =0$ and 
\begin{equation}\label{e*}
e^* =\frac{ \left ( 1 - \alpha \right ) \left ( r -1  \right )-\alpha \gamma} {\left ( 1 - \alpha \right ) \gamma },
\end{equation}
provided that $\alpha >  \left ( r -1  \right )/\left ( \left ( r -1  \right ) - \gamma \right )$, which guarantees that $e^* < 1$. Otherwise, we have $e^* =  \left ( r -1  \right )/\left (\delta + \gamma \right ) > 1$ (see Section \ref{subsec:FE}). Figure \ref{fig:2} shows the finite-engineers fixed points at the transition, where the  stable and unstable segments are flagged with different colors. The  figure reveals that the fixed point $e^* >0$  is stable  for small $\delta$, as expected, and that it is also stable close to the point $(\delta_c, \alpha_c)$. This indicates that this point does not belong to the drop-shaped curve shown in Fig.\ \ref{fig:1}, which  delimits the regions of stability of the finite-engineers fixed point. This facet will become much more evident when
we consider larger values  of the recovery probability $\rho$.

Another interesting  feature of Fig.\  \ref{fig:1} is the appearance of a region of bistability of the two fixed points for small values of $\delta $ and $ \alpha < \delta/3 $. The dependence on the initial density of engineers, as well as on the  initial habitats fractions, that characterizes the bistability regime results in an Allee  effect where
  the environment cannot sustain a low density of engineers, as the change of the virgin habitats into usable ones  can only be achieved by high-density populations.

Figure \ref{fig:3} illustrates the  time dependence of the density of engineers $e_t$ and of the fraction of virgins habitats  $v_t$ for three representative regions of  Fig.\  \ref{fig:1}, where the zero-engineers fixed point ($e^* =0$) is unstable, i.e, for $\delta < \alpha \left ( r -1 \right)$, so the results do not depend on the initial conditions. The time dependence of the fraction of usable habitats $h_t$ is  similar to that of $e_t$.  The recursion equations (\ref{e})-(\ref{v})  were iterated using quadruple precision. The oscillatory convergence to the finite-engineers fixed point ($e^* >0$) is the effect of the complex eigenvalues of the Jacobian matrix (\ref{Je*}). These oscillations may  lead to extremely low values of the  density engineers (e.g., $e_t$ on the order of $10^{-10}$), hence the need of  a high precision numerical scheme to iterate the recursion equations. Since $e^* =0$ is unstable, the population  eventually recovers from these near-extinction events as illustrated in the middle and right panels of Fig.\  \ref{fig:3}.

The nature of the transition between the periodic solutions and the zero-engineers fixed point is clarified in Fig.\ \ref{fig:4} where we fix the decay rate per generation to $\delta=0.15$ and  show  $e_t$ and $v_t$  for different values of  $\alpha$ in distinct panels. The left and middle panels, which show results for values of $\alpha$ very close to the stability boundary  $\alpha = 0.05$, reveal what happens: the period of the oscillations increases as that boundary is approached and diverges when the zero-engineers fixed point becomes stable, so that $e^* =0$ and $v^*=1$ for all times.
The right panel in Fig.\ \ref{fig:4} reminds us  that the periodic solutions are not necessarily associated to  near-extinction events and that they can reproduce  smooth oscillations with the usual phase shift of one-quarter of a cycle between engineers and virgin habitats, which characterizes the  predator-prey  models  \cite{Turchin_03}.

\begin{figure*}
\begin{center}
 \subfigure{\includegraphics[width=0.325\textwidth]{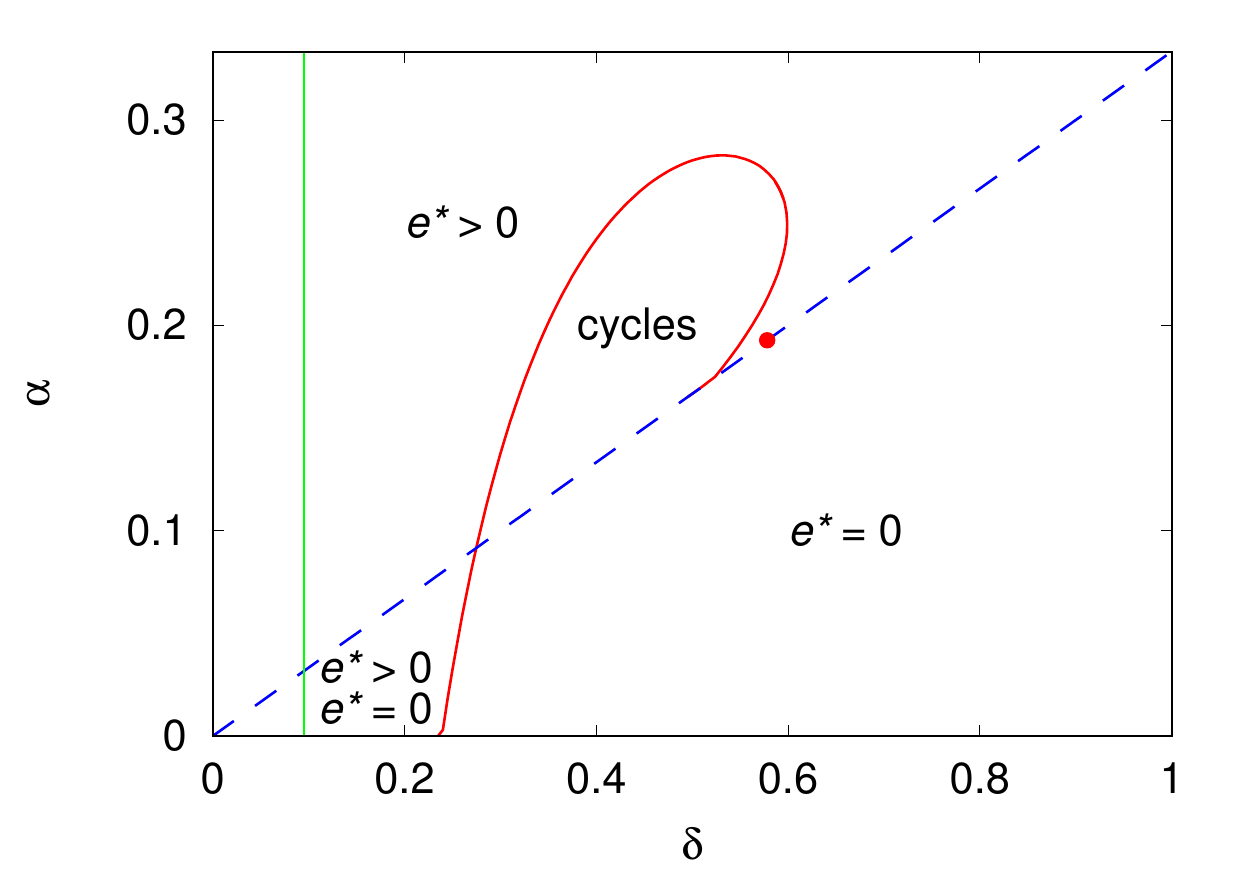}} 
 \subfigure{\includegraphics[width=0.325\textwidth]{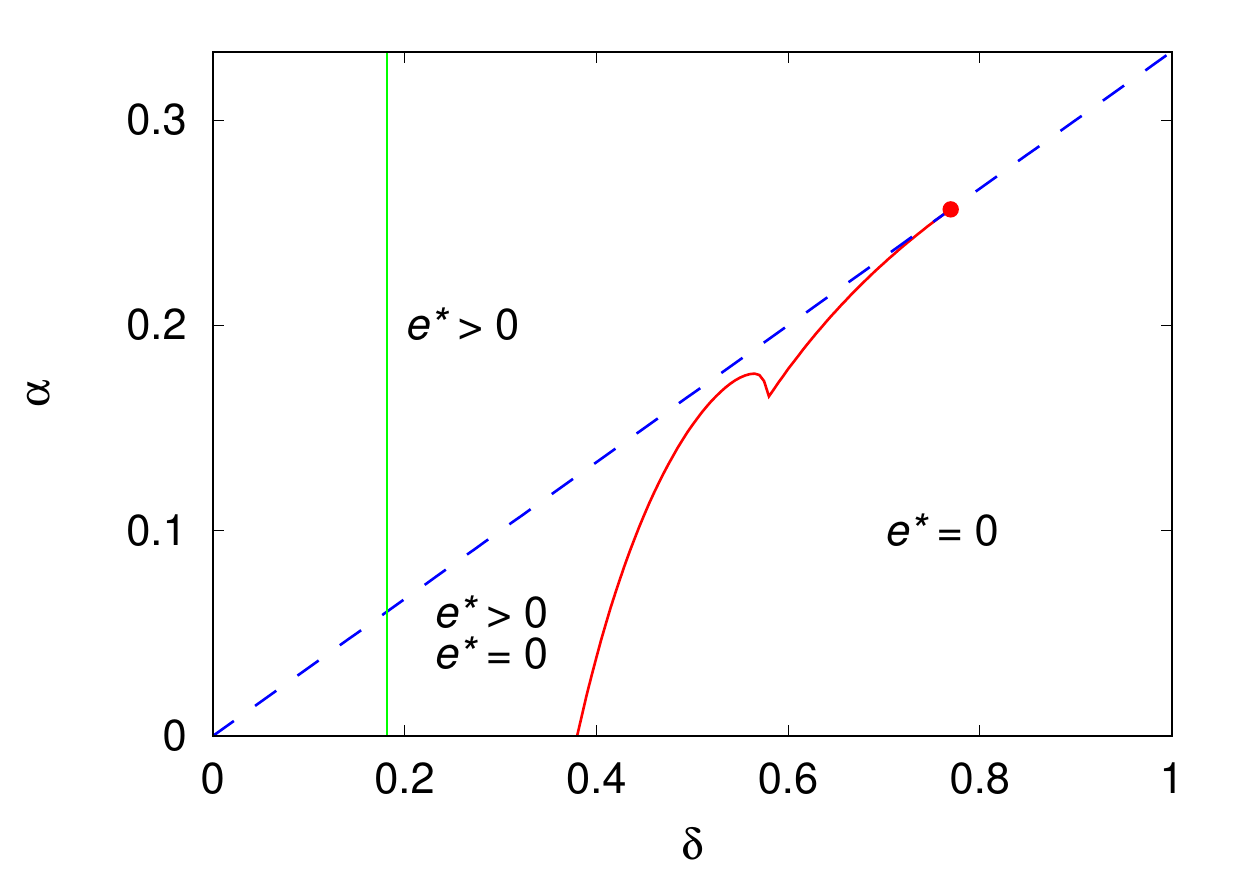}} 
 \subfigure{\includegraphics[width=0.325\textwidth]{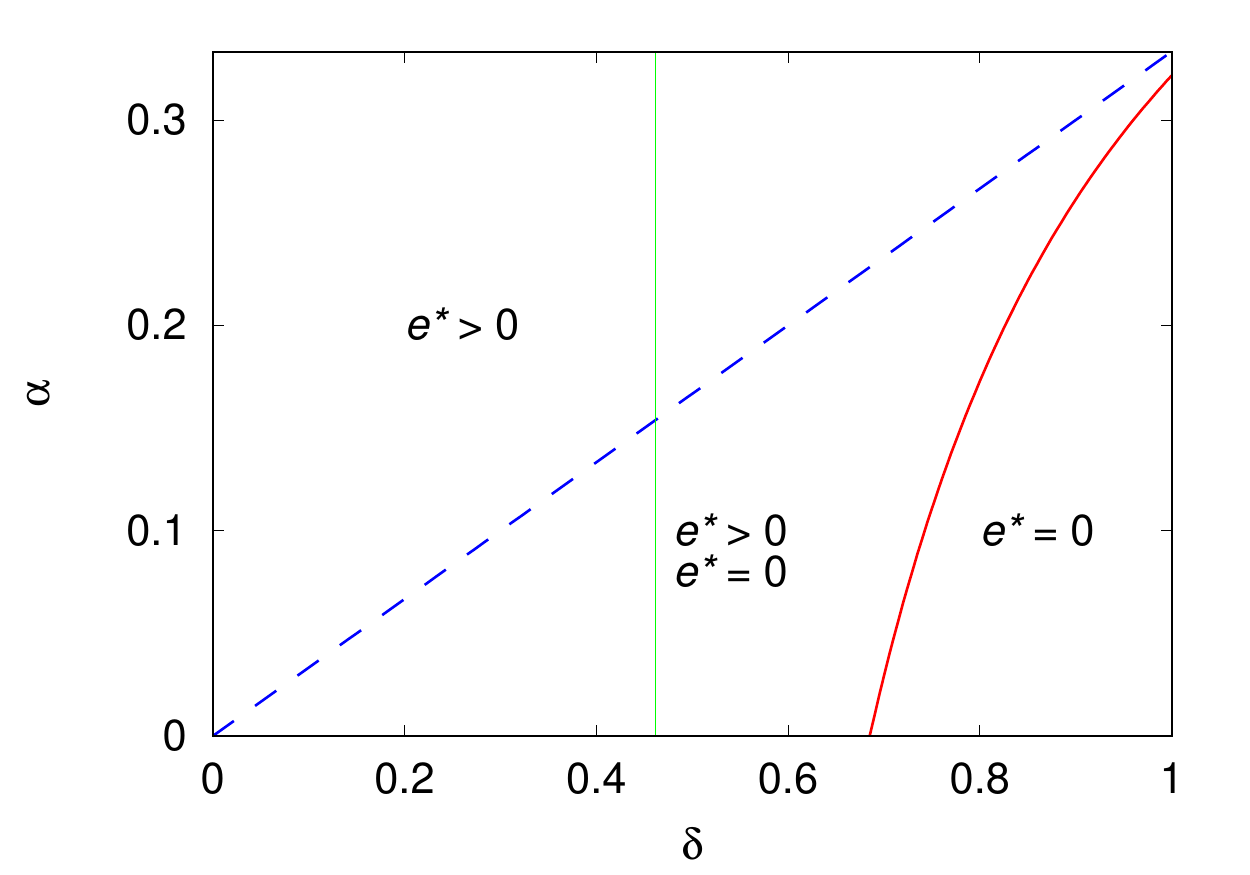}}
\caption{Regions in the space of the parameters  $\left ( \delta, \alpha \right )$ where the  fixed point  solutions are locally stable for   $r=4$ and   $\rho =0.05$ (left panel), $\rho =0.1$ (middle panel)  and $\rho =0.3$  (right panel). The zero-engineers fixed point is stable below the dashed blue  line $\alpha = \delta/3$. The fixed point $e^* > 1$ exists and is stable only in the regions to the left of the vertical  green lines. The bullets  at the dashed line indicate the values of $\delta$  above which the transition between the finite and the zero-engineers fixed points is continuous.    }
\label{fig:6}
\end{center}
\end{figure*}

Next we consider the effects of varying the parameters $r$ and $\rho$ that were kept fixed in the above  analysis.
Figure \ref{fig:5} shows the regions of stability of the fixed points for three  values of the  intrinsic growth rate  $r$. To keep the size of the region where the zero-engineers fixed point is stable we use the scaled variable $\delta/\left ( r-1 \right )$ as the independent variable. The results indicate that increase of $r$ increases the size of the regions of limit cycles  and decreases the size of the  regions of  bistability, so variation of $r$ produces quantitative effects  only. This is in stark contrast with the effects of varying   the regeneration rate per generation $\rho$ shown in Fig.\ \ref{fig:6}.  In fact,  increase of $\rho$ decreases the size of the region of limit cycles, which eventually disappears altogether for large $\rho$, since there is no region left  in the space of parameters where the fixed points $e^* > 0$ and $e^* = 0$ are simultaneously unstable.  For instance, the curve for  $\rho = 0.3$ shown in the right panel of Fig.\ \ref{fig:6} is given by the condition that the discriminant of the quadratic equation (\ref{e*2})  is zero,  so for large $\rho$  the finite-engineers fixed point is locally stable wherever it is real.  The  piecewise curve for 
$\rho =0.1$ is given by the composition of the conditions that $e^*$ is real and that the norms of the eigenvalues of the Jacobian (\ref{Je*}) are less than 1.
In addition, increase of $\rho$ greatly increases the region of bistability as well as the region  where $e^* > 1$.

%----------------------------------------------------------------------------------
%----------------------------------------------------------------------------------
 
\begin{figure}
\begin{center}
\includegraphics[width=0.48\textwidth]{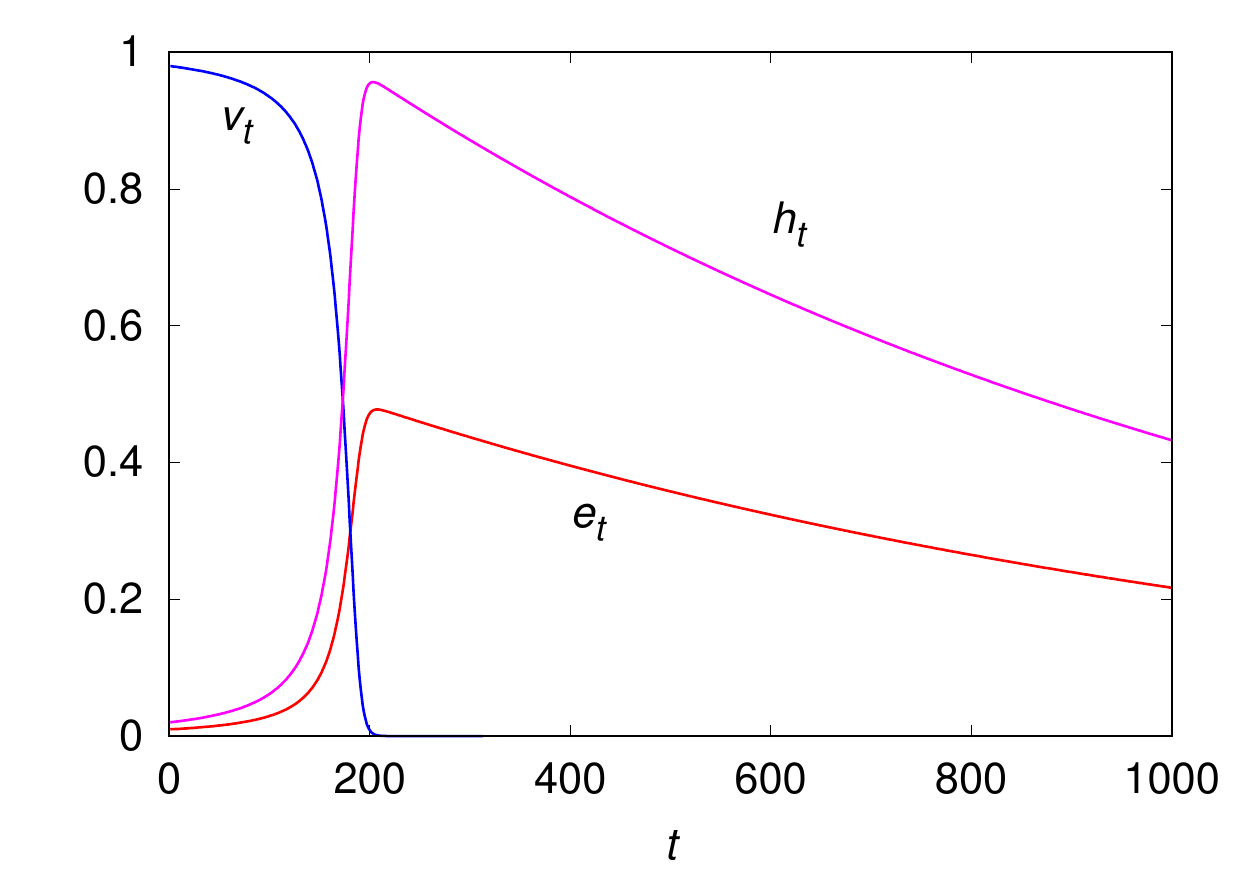}
\caption{Density of engineers $e_t$ (red curves),  fraction of virgin habitats $v_t$ (blue curves)  and  fraction of usable habitats $h_t$ (magenta curves) for $r=1.5$, $\rho = 0$, $\delta =0.001$ and $\alpha = 0.01$.  The dynamics is attracted to the doomsday  fixed point $v^*=e^* = h^*= 0$ and $d^* =1$. }
\label{fig:7}
\end{center}
\end{figure}

To conclude, we consider now a simple but instructive scenario where the natural resources are not renewable, i.e., $\rho =0$. Figure \ref{fig:7} shows the time evolution of the population towards the doomsday  fixed point $v^*=e^* = h^*= 0$ and $d^* =1$ where only the degraded habitats are left. We have not considered this fixed point in Section  \ref{sec:zero} because it exists and is stable solely for $\rho=0$ and  $\delta >0$, regardless of the values the other model parameters. The interesting aspect this figure highlights  is that the population survives for a long time after the irreversible disappearance of the virgin habitats, thus revealing that the usable habitats  introduce a delay on the  influence of the  virgin habitats on the population that is absent in the usual predator-prey models. However, this delay appears in more sophisticated models of human-nature interactions in which humans accumulate surpluses (i.e., wealth)  and use them when natural 
resources are   scarce or  unavailable \cite{Motesharrei_14}.

%----------------------------------------------------------------------------------
%----------------------------------------------------------------------------------

\section{Discussion}\label{sec:disc}

The derivation of the discrete-time recursion equations  (\ref{e})-(\ref{c}) assumes non-overlapping generations,  which is clearly not the case for human populations. Our justification to use this  approach  is  purely pragmatic since it is much easier and much less error-prone to iterate those recursions than to solve numerically the differential equations of a continuous-time model, particularly in the region of reversible collapses where the density of engineers  is very close to zero and the periods of the limit cycles are exceedingly long. However, as the  solutions do not exhibit short-time fluctuations and are all smooth in the scale of a few generations we do not expect any significant differences between the discrete and the continuous time approaches. In addition, 
the discrete-time approach can easily be  extended to describe space-dependent problems, resulting in coupled map lattices  that can be thoroughly studied numerically. In particular, in this context a different type of population collapse can be observed when the expanding colonies of engineers reach the boundaries of their territories, thus ending the exploration of 
new virgin habitats and resulting in a sharp drop on the population density   \cite{Fontanari_18}. 

In agreement with the findings for the predator-prey type models \cite{Brander_98,Motesharrei_14}, we find that a key parameter to determine the long-term outcome of the population dynamics  of ecosystem engineers is the regeneration rate per generation  $\rho$, which determines 
the renewability of the  resource base (see Fig.\ \ref{fig:6}). Cycles of prosperity, collapse and revival occur in the case the degraded habitats have a slow regeneration rate. In the extreme case of non-renewable resources (i.e., $\rho=0$), the outcome of the dynamics is the irreversible collapse of the entire ecosystem (see Fig.\ \ref{fig:7}). In the other extreme, when  the degraded habitats rapidly recover into virgin habitats, the limit cycles disappear altogether and the outcome of the dynamics is determined by the basins of attraction of the finite and zero-engineers fixed points.

The local stability analysis of the zero-engineers fixed point yields a simple condition for the survival of the population, viz., the product of the parameter that measures the effective population growth rate (i.e., $r-1$) and  the parameter that measures the efficiency of the transformation of virgin into usable habitats (i.e., $\alpha$)  must be greater than the decay rate per generation of the usable habitats, i.e., $ \alpha \left ( r -1 \right ) > \delta$.  
Hence  increase of  the efficacy of the engineers to explore the virgin habitats or increase of  their intrinsic effective growth rate will both guarantee  the  survival of the population,  regardless of the regeneration rate $\rho > 0$ of the degraded habitats. The characteristics of the steady-states 
for large $r$ and large $\alpha$, however, are completely distinct.

On the one hand, increase of $r$ favors  limit cycles characterized by  long periods of  time  when $e_t \ll 1$, which we refer to as reversible collapses.  These collapses  are reversible because they occur in  regions of the model parameters where the zero-engineers fixed point is unstable.  This scenario is expected since a large value of $r$ produces a large density of engineers, which, according to the prescription (\ref{c}), can transform all virgin habitats in a single generation, thus producing  the cyclical pattern of feast and famine   illustrated in the left and middle panels of Fig.\ \ref{fig:4}. 

On the other hand, increase of $\alpha$ favors the finite-engineers fixed points and leads to a  stable   balance between population and resources. In our setup, a small $\alpha$  corresponds to a  low-tech society where the exploration of the virgin habitats requires a high density of individuals. Interestingly, our model predicts that technology improvements that allow a small population to explore efficiently  the available natural resources, a situation that is modeled by increasing $\alpha$ while keeping all the other parameters fixed, can eliminate the dangerous cycles of prosperity, collapse and revival.   

This result is especially notable because it neatly expresses  the argument put forward by the growth optimists, namely,  that technological progress  can reduce the impact of  resource depletion  and avoid environmental and economic collapse \cite{Atkisson_10}.  Of course, this happens because our model does not account (among many other things) for the Jevons effect that
occurs when technological progress  increases the efficiency with which a resource is extracted, but the rate of consumption of that resource rises due to increasing demand \cite{Polimeni_15}. In our model, increasing demand  amounts to considering the decay rate per generation dependent on the availability of  usable habitats, i.e., $\delta = \delta \left ( h_t \right )$, so  that the more goods are available, the higher  their consumption by the population. The possibility of supporting the growth optimists viewpoints as well as addressing quantitatively the Jevons effect, the analysis of which we  postpone for a future contribution,  highlights the potential of the ecosystem engineers approach to model both ecological and economic aspects of the human-nature dynamics.

\bigskip

\acknowledgments
J.F.F. is  supported in part by Grant No.\  2017/23288-0, Fun\-da\-\c{c}\~ao de Amparo \`a Pesquisa do Estado de S\~ao Paulo 
(FAPESP) and  by Grant No.\ 305058/2017-7, Conselho Nacional de Desenvolvimento 
Cient\'{\i}\-fi\-co e Tecnol\'ogico (CNPq). GML is  supported by a CAPES scholarship.

%\section*{Supplementary (if necessary)}

\end{document}